%% file: main.tex
\documentclass[twocolumn,12pt]{article}
\usepackage[utf8]{inputenc}
\usepackage{setspace} 
\usepackage{xpatch}

\title{ {\large \bf An efficient deception architecture for
cloud-based virtual networks} }

\author{
	\begin{tabular}{c}
	Mohammed Qasem,  Hussain M. J. Almohri* \\
	\end{tabular}\\
	{\it Department of Computer Science}
	\\ {\it Kuwait University, Kuwait}
   \\ {\it *Corresponding author:  almohri@cs.ku.edu.kw}
}

\usepackage[table,xcdraw]{xcolor}
\usepackage{pgf, tikz}
\usetikzlibrary{arrows, automata}
\usepackage{graphicx}

\usepackage{fancyhdr}
\usepackage{lastpage}
\usepackage{pgfplots,multicol}

\usepackage{times}
\usepackage[a4paper,margin=25mm]{geometry}

\usepackage[american]{babel}

\usepackage[bf,figurename=Fig.,labelsep=period]{caption}

\usepackage{secdot}
\usepackage[compact]{titlesec}
\titleformat{\subsection}
{\normalfont\normalsize}{\thesubsection}{1pt}{~}
\titleformat{\subsubsection}
{\normalfont\normalsize}{\thesubsubsection}{1pt}{~}
\titlespacing{\subsection}{0pt}{0.4cm}{0pt}
\titlespacing{\subsubsection}{0pt}{0.4cm}{0pt}

\titleformat{\section}{\bf}{\thesection}{1pt}{.~}

\begin{document}

\date{ }

\twocolumn[
  \begin{@twocolumnfalse}
    \maketitle
    \thispagestyle{empty}
   \begin{center}
    {\bf Abstract}
    \end{center}
    \noindent Emerging deceptive systems present a new promise for the uprising security problems in cloud-based virtual networks, especially those operated by small and medium enterprises. The main goal of deceptive systems is to form a layer of defensive nodes in an Internet-accessible cloud-based virtual network to distract and deceive malicious clients. While numerous approaches provide distinct models for developing decisive systems,  misery digraphs present a promising decisive model for distracting powerful remote intrusions. Misery digraphs can delay access to targets deep in a cloud-based virtual network. A central challenge to the theory of misery digraphs is verifying their applicability in prominent cloud computing platforms as well as measuring the efficiency of networks that adapt them. Thus, architectural support is needed that can be realized with long-term support technologies and can be deployed for large networks. This work presents and analyzes a high-throughput architecture for misery digraphs, embarking on implementation details and performance analysis. Full implementation of the architecture in Amazon Web Services imposes modest performance delays in the request processing, while highly delaying stealth intrusions in the network.\newline\newline
	\vspace{1cm}
  \end{@twocolumnfalse}
]

\input{1}
\input{2}
\input{3}
\input{4}
\input{5}
\input{6}

 \bibliographystyle{abbrv}
\bibliography{references}

\end{document}

%% file: 1.tex
\section{Introduction} Cloud-based virtual networks enable small and medium
enterprises to rapidly and efficiently initialize, deploy, maintain, and evolve
networks of virtual machines. For example, a startup company can utilize a
virtual network for connecting a user-end client application to the company's
services by launching virtual machine instances that could be conveniently
connected to Internet gateways, subject to a firewall, and access control rules.
Prominent cloud computing platforms such as Amazon Web Services and Google
Compute Engine provide both programmable interfaces to manage virtual machine
instances and modify their access control rules. While solutions have assisted
small and medium enterprises in achieving their rapid growth, security
challenges continue to threaten these networks, causing unprecedented costs as a
result of attacks, which can lead to disastrous consequences. 

Among the many security problems, remote vulnerabilities are of critical
importance. A remote vulnerability in an Internet service can potentially allow
intrusion into hosts that constitute a network's surface. These hosts are
connected to the Internet, receive requests from clients, and communicate them
to isolated databases, application servers, and other services within the
network. The intrusion's goal is to gain remote execution access on the victim
host, for example, through opening a remote shell, or by hijacking a vulnerable
application to divert the execution, ultimately creating malicious user accounts
on the target. When under control of the intruder, the victim host in the
network's surface would allow the attacker to further exploit vulnerabilities
within the network by investigating hosts that are accessible from the
compromised one. 

Network deception is a potential solution for slowing the rapid progress of
intrusion in a virtual network. To this extend, in previous work, the concept of
misery digraphs was introduced, which provided a dynamic structure within a
network of web services, distracting and confusing the attacker who wishes to
compromise a specific target deep in the network~\cite{8125739}. In the proposed
solution, a cloud-based virtual network is modeled as a connectivity digraph
representing the network's accessibility structure. The connectivity digraph is
then converted into an expanded structure of decoy nodes that are dynamically
modified over time, consistently losing an intruder's effort towards a target. 
Misery digraph is a powerful concept that requires intensive implementation and
testing. 

The present work investigates an architecture of the misery digraphs that
enables a dynamic network structure within Amazon Web Services. The architecture
includes several components that, given an initial network setup, transform an
existing cloud-based virtual network into one that includes misery digraphs,
implementing a full misery digraph defense system. This requires a careful
design of a transformation process, a realistic implementation of misery
digraphs using a service-oriented network, and an analysis of the feasibility of
the proposed system.

\subsection{Problem statement} This work investigates the problem of the
efficient transformation of web services in cloud-based virtual networks into
deceiving networks containing misery digraphs. The focus is on networks that are
created on Amazon Web Services (AWS) with complex structures containing multiple
web servers, application servers, and database servers. The assumption is that
the web servers, which form the network's surface, are vulnerable to remote
attacks, and the attacker does not have prior access to servers. The attacker's
target is a critical asset, such as a database server in the network. 

\subsection{Approach and results} The approach is to design an architecture that
realizes misery digraphs in AWS. Since misery digraphs complicate an attack path
(by enlarging the path to the target, adding decoys that are continuously
relocated), the challenge is to minimize the performance penalty facing benign
network requests. Thus, we designed a transformation process that aids network
managers to implement misery digraphs according to the specifications
in~\cite{8125739}. The transformation process receives a conventional (and
vulnerable) virtual network and produces a network of misery digraphs, including
the network paths of the original conventional network, while adding decoy
paths. Misery digraphs evolve and change their structures over time, increasing
the system's uncertainty for attackers. This dynamic nature of misery digraphs
requires a special proxy system for forwarding network requests from the
network's entry points towards the target. We developed the proxy system for
misery digraphs based on Apache's reverse proxy module. The efficiency of the
approach was measured by constructing two networks that perform identical
functions, one using misery digraphs as the underlying topology, and one that
uses a minimal connectivity digraph. The results show that misery digraphs
impose a modest performance penalty when processing HTTP requests when compared
to networks that do not implement misery digraphs.  

%% file: 2.tex
\section{Background} While no prior work has proposed a practical and efficient
architecture for deceiving systems for the cloud, the literature includes a wide
spectrum of deception and moving target defense techniques for combating
powerful network attacks.  Some of the earlier pioneering works in deception
focused on the use of overlay networks as the core idea of deceiving,
distracting, and slowing denial of service attackers targeting specific hosts
within a network.  Secure Overlay Services (SOS)~\cite{1258124}, and later
WebSOS~ \cite{Morein:2003}, utilize an overlay network and enforce strict
verification of the sources of incoming requests when communicating with a host.
If a source passed verification, a subset of hosts would act as proxies that
forward the traffic towards hidden servers (often serving applications) within
the network. The proxies are secret, and their identities are not exposed. The
assumption in SOS and WebSOS is that both parts of the communication are known a
priori. 

Network overlays, the use of proxy servers, and the assumption of known clients
was the underlying approach for many other related works.  Within this context,
Migrating OVErlay (MOVE)~\cite{tavrouKNMR05} was introduced, advocating the idea
of client filtration via authentication, and used client migration as a policy
for maintaining service availability as well as detecting abusing clients (also
called insiders). MOVE, similar to others, built on ideas from moving target
defenses~\cite{7122306,Evans2011}. The idea behind MOTAG~ \cite{6614155} was to
provide a hidden contact point to each legitimate client when the client is
registered and authorized to use the service. MOTAG uses these hidden contact
points (or proxies) to filter clients and control access to application servers.
Later, address shuffling and client migration are also used in MOTAG to bypass
attacks. Moving target defenses do not necessarily aim to deceive attackers, but
to keep attackers in the dark, random port hopping was proposed~\cite{4288181},
which distracts denial of service attackers while using packet filtration to
recognize legitimate traffic. Similar to randomizing ports, redundant data
routing paths~\cite{Lee:2007,5432179} is a technique that can potentially
distract attackers.

Defending against denial of service in clouds has been the subject of some other
recent studies. One proposed approach is to use elastic cloud features to guard
against a growing distributed denial of service attack~\cite{6903585}. Misery
digraphs avoid the ever-expanding networks due to denial of service
vulnerabilities by modifying the existing network of machines, thus preventing
overhead costs.

The idea of deception has also been widely studied in various other forms,
including deceptive attack
techniques~\cite{Spitzner:2003,Lisy:2010,Alowibdi:2014}, defense techniques that
use software defined networks to deceive
attackers~\cite{Jafarian:2012,Achleitner:2016}, defending against non-volumetric
distributed denial of service attacks~\cite{Pal:2017}, slowing down network
scanners~\cite{Alt:2014}, occasional trap-setting to detect illegitimate insider
activities~\cite{Bowen:2010}, and as virtualized honeypots atop the production
network~\cite{Stoecklin:2018,Han:2017}.

\subsection{Misery Digraphs}\label{sec:mdg} Misery digraphs~\cite{8125739} take
a radical approach and form a the theoretical basis for deception in cloud-based
virtual networks.  Similar to attack
graphs~\cite{7122306,Miehling:2015,7056435}, misery digraphs model host access
control rules in a cloud computing platform as a digraph. The resulting digraph
is input to an algorithm to enlarge and stretch every path from an entry point
to a target host. An entry point is a host that is accessible over the Internet
without filtration based on origin's IP, while the target is a host that is only
accessible through entry points or other hosts that are accessible through entry
points. 

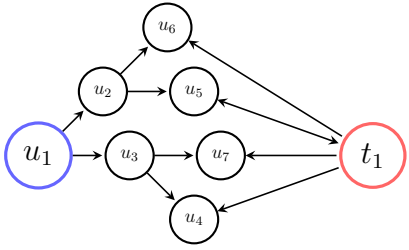
\begin{figure}
\centering
    \begin{tikzpicture}[
            > = stealth, 
            shorten > = 1pt, 
            auto,
            node distance = 2cm, 
            semithick, 
            targetnode/.style={circle, draw=red!60, fill=white!5, very thick, minimum size=2mm},
            entrynode/.style={circle, draw=blue!60, fill=white!5, very thick, minimum size=2mm},
            scale=0.6,
        ]

        \tikzstyle{every state}=[
            draw = black,
            thick,
            fill = white,
            scale=0.6
        ]

        \node[entrynode] (u1) {$u_1$};
        \node[state] (u2) [above right of=u1] {$u_2$};
        \node[state] (u5) [right of=u2] {$u_5$};
        \node[state] (u6) [above right of=u2] {$u_6$};
        \node[state] (u3) [right of=u1] {$u_3$};
        \node[state] (u4) [below right of=u3] {$u_4$};
        \node[state] (u7) [right of=u3] {$u_7$};
        \node[targetnode] (t) [right of=u7] {$t_1$};

        \path[->] (u1) edge node {} (u2);
        \path[->] (u1) edge node {} (u3);
        \path[->] (u3) edge node {} (u4);
        \path[->] (u3) edge node {} (u7);
        \path[->] (u2) edge node {} (u5);
        \path[->] (u2) edge node {} (u6);
        \path[<-] (u7) edge node {} (t);
        \path[<-] (u6) edge node {} (t);
        \path[<-] (u4) edge node {} (t);
        \path[<->] (u5) edge node {} (t);

    \end{tikzpicture}
\caption{A misery digraph generated for the attack path $ u_1 \rightarrow u_2 \rightarrow t_1 $, with an embedded binary tree. An edge $u_i \rightarrow u_j$ means $u_i$ can directly access $u_j$ via some network protocol. In reality, $u_1$ is a web server, $u_2$ and $u_3$ are application servers, $t_1$ is a database server, and the rest are reverse proxies. }
  \label{fig:md1}
\end{figure}

The core elements of misery digraphs are: (1) multiple, identical, and enlarged
paths to a target, and (2) a schedule of resetting and relocating hosts on
randomly selected paths to target. For example, a simple path to target, $$ u_1
\rightarrow u_2 \rightarrow t_1 $$ is converted into a digraph consisting of a
$k$-ary tree, with a single {\it enabled} path towards $t_1$, as depicted
in~\ref{fig:md1}.

A cloud-based virtual network implementing misery digraphs would need to
replicate a network request to nodes directly accessible from an entry point.
For example, in Figure~\ref{fig:md1}, a request $R_1$ received at $u_1$ will be
replicated and sent to $u_2$ and $u_3$, which in turn replicate $R_1$ to the
layer below. The nodes at the final layer $u_4$, $u_5$, $u_6$, and $u_7$ attempt
to forward $R_1$ to $t_1$. Misery digraphs guarantee that $R_1$ will reach $t_1$
through one and only one path (through $ t_1 \rightarrow  u_7$ in the example
digraph of Figure~\ref{fig:md1}). The rationale is to confuse the attacker early
on in the digraph about the true path towards the target. 

Misery digraphs are also required to maintain freshness through a schedule of
changes to the location of nodes. At fixed time intervals, a controller
procedure inside the network selects two nodes at the same layer (for example,
$u_4$ and $u_6$ in Figure~\ref{fig:md1}) and switches their locations in the
digraph. Next, the hosts representing the nodes are deleted from the network,
and two fresh hosts are created and added in their location. The hosts will be
created from images that contain the required software for processing requests. 

While simulation is a useful method for assessing the effectiveness of misery
digraphs and similar models, this work attempts to provide a realizable and
practical architecture, addressing challenges facing system administrators when
adapting the model, including challenges concerning the required implementation
and performance tuning. 

%% file: 3.tex
\section{Architecture}

\begin{figure}[tb] \centering \includegraphics[scale=0.26]{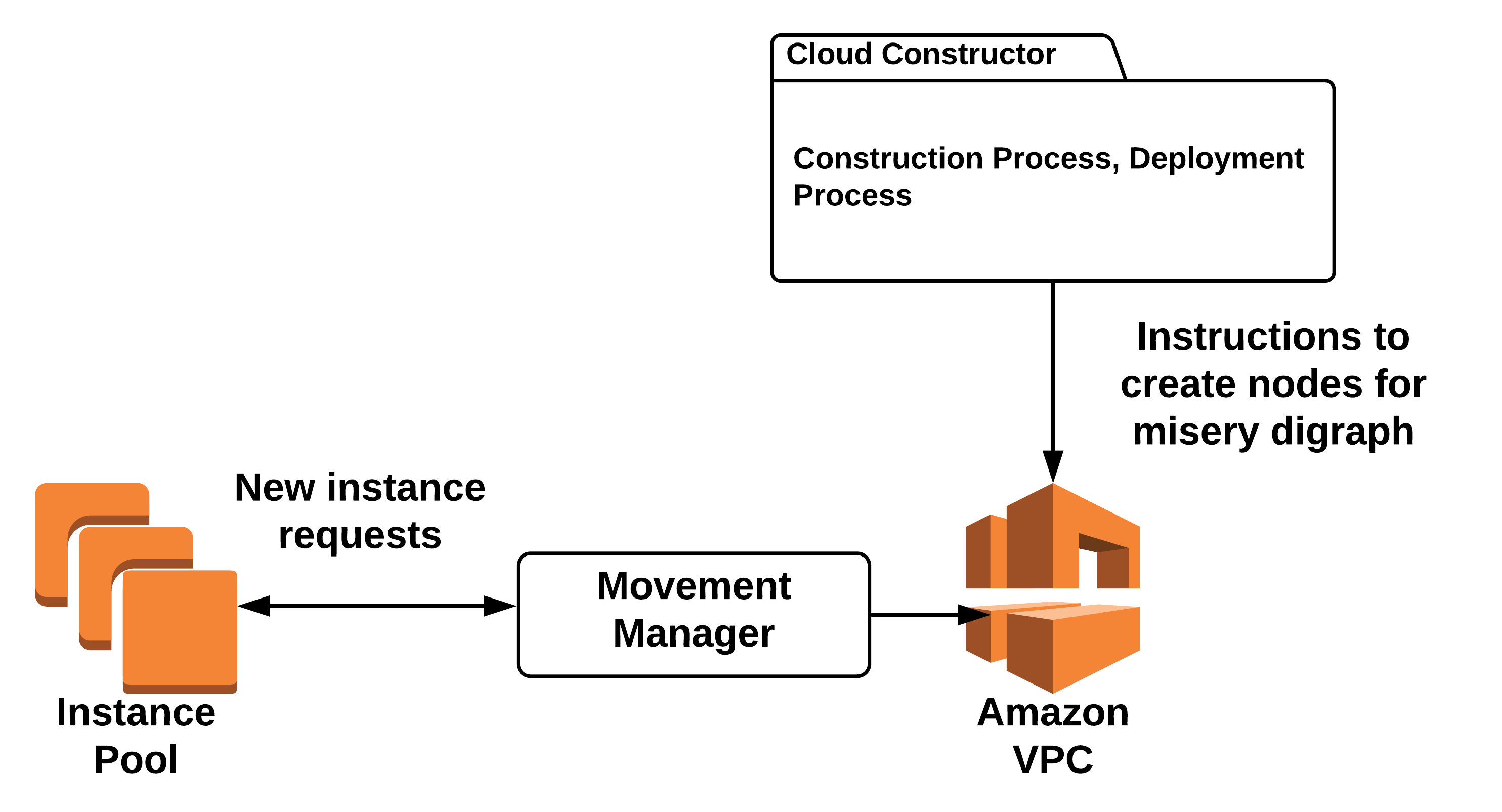}
\caption{Our architecture includes a cloud constructor, which examines the
initial topology of the cloud-based virtual network, expands it to one with an
embedded misery digraph of decoy nodes with a special access control setting,
and automatically deploys the resulting topology in the network's Virtual
Private Cloud (VPC). A movement manager uses a pool of initialized virtual
machines to continuously evolve the resulting network by a systematic mutation
of the misery digraph. } \label{fig:MTDComponentOfMDGImplementation}
\end{figure}

Our design of misery digraphs has two prime components. The first is a component
to implement the continuous evolution of the digraph
(Figure~\ref{fig:MTDComponentOfMDGImplementation}) resulting in a Moving Target
Defense (MTD). The second component implements a deep traversal of the client's
request through the Misery Digraph Cloud, sending the response back to the
client. This process follows a multicasting method, as described in
Section~\ref{sec:mdg}.  The focus of the architecture is on implementing two
tasks:

\begin{enumerate} \item Constructing a Misery Digraph Network using pre-existing
cloud architecture and deploying the constructed network to the cloud as a set
of instances and firewall rules (security groups rules in AWS terms).  \item
Frequently selecting random instances of Misery Digraph Cloud to switch their
parents and children and reset their images.  This process is referred to as the
\textit{Transformation Process} (TP).  \end{enumerate}

Throughout this work, \textit{Misery Digraph Network} refers to the theoretical
representation of the network as a graph data structure, and \textit{Misery
Digraph Cloud} refers to the realization of misery digraphs in the cloud as a
set of instances and security groups. Also, a client is defined as an
application that is served by the Misery Digraph Cloud.

Figure \ref{fig:MTDComponentOfMDGImplementation} shows the Moving Target Defense
component of our architecture. As a first step, the administrator builds the
architecture of the cloud and defines firewall rules, then \textit{Cloud
Constructor} creates a Misery Digraph Network using the same method as in
\cite{8125739}, filling the network with decoys and deploy this network to the
cloud.  Once the Misery Digraph Cloud is ready, \textit{Movement Manager}
selects two random instances from a layer in the misery digraph, which only
contains sibling nodes. The first layer and the last layer (layer at depth
$d+1$, containing the target node) are excluded from selection by the Movement
Manager.  Note that a misery digraph is created using two parameters: a
branching factor $k$, and the number of layers $d+1$.  Once the layer is
selected, the Transformation Process is performed at the selected layer. This
entire process of selection and transformation is repeated each period of time,
$t$.

\begin{figure}[h]
\includegraphics[scale=0.3]{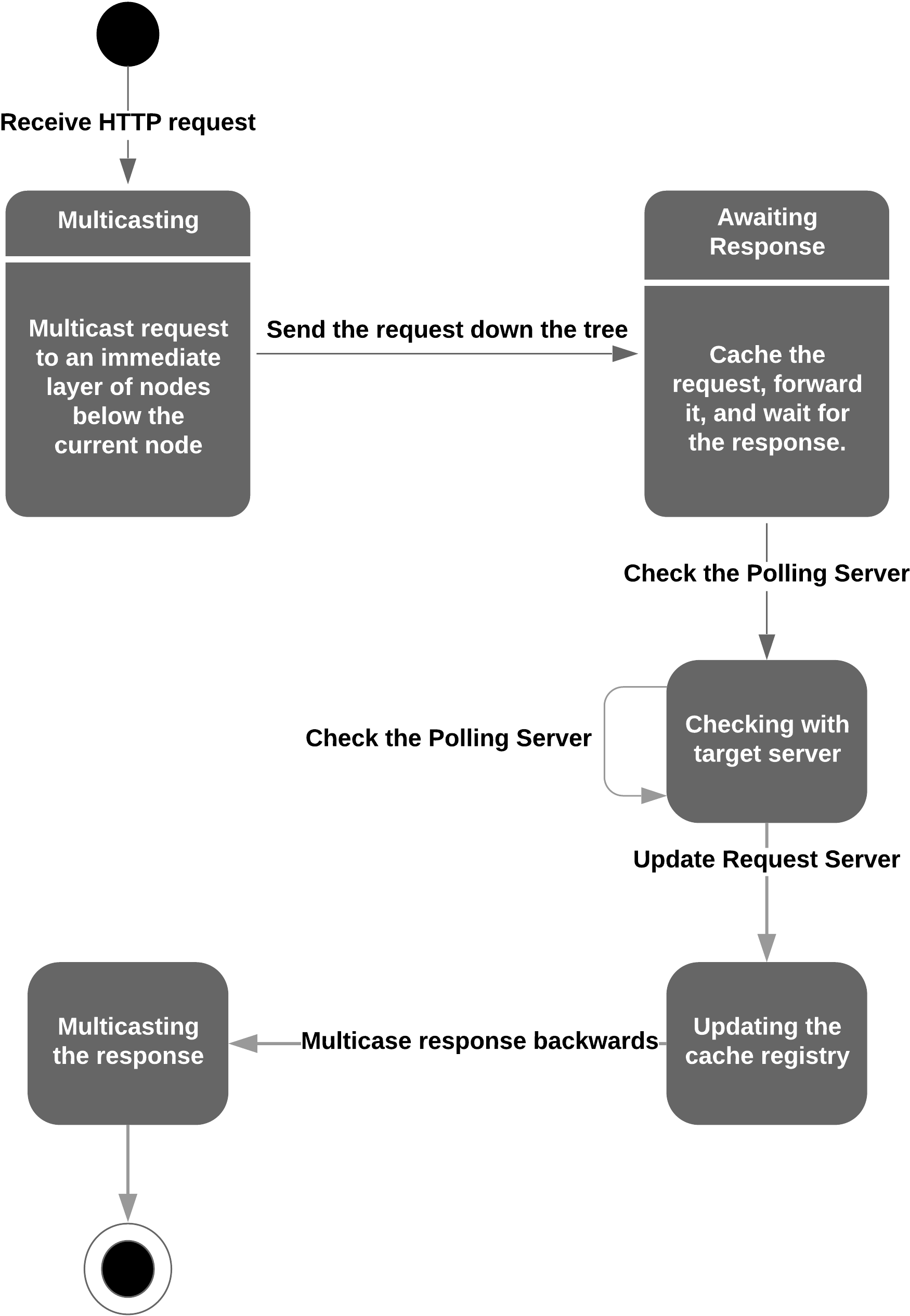}
\centering
    \caption{Life cycle of a client's request in the Misery Digraph Cloud.}
    \label{fig:ClientRequestComponentOfMDGImplementation} \end{figure}

Figure \ref{fig:ClientRequestComponentOfMDGImplementation} shows the life cycle
of client's request in the MDG cloud. In our architecture, the life cycle of a
request starts by receiving it in an entry point node. Compared to normal
processing of an HTTP request, which is sent to an application server and
finally to a database server, our architecture modifies this path by
multicasting the request to a layer of decoy nodes. This multicasting continues
until it reaches a Request Server instance from which a database request is
created. The requests are cached in a database registry, awaiting responses from
the database server. Once the response is received, it is propagated back up the
tree until it reaches the entry point. 

The entry point instance is the only instance of Misery Digraph Cloud that is
publicly accessible. A client sends a request to the entry point, which runs the
\textit{Misery Multicaster} that multicasts the requests received from parent
nodes in the previous layer to all children in the next layer.  Once the request
reaches layer $d$, a request to the target is stored by the \textit{Requests
Server}. The RS will then wait for the PS to ask about the request. The Polling
Server runs on the target itself and connects to Requests Servers on layer $d$
to query them if there is any new request for the database. If so, the requests
will be processed, and the response will be sent to the Requests Servers. A
response travels backward (in the opposite direction of the leaves) until it
reaches the client through the entry point.

\textit{Isolated Target} refers to disallowing any instance of layer $d$ to
reach the \textit{target}, letting the target to poll the requests instead. This
is one of the differences from what is presented in the original work
\cite{8125739}.

The presented architecture is intended for testing on Amazon Web Services (AWS).
Despite this, the main ideas are applicable to competitors because the
architecture was designed to be generic with the least dependency on the
underlying technology specific to AWS. Throughout this work, references are made
to AWS Elastic Compute Cloud (EC2) instances, which are virtual machines that
are computationally independent and are created using programming tools
available to cloud users. A cloud refers to an account on a cloud computing
platform containing a virtual network of EC2 instances. 

\subsection{Cloud constructor}

In order to construct the Misery Digraph Cloud, the cloud the administrator
should set the preferred values of the parameters $d$ and $k$, the parameter $d
+ 1$ represents the number of the misery Digraph Network's layers, while $k$
represents the number of children for each node in the graph that belongs to
layer $1,\ldots,d$.  After setting the parameters $d$ and $k$, the Cloud
Constructor uses the Amazon Web Services Application Programming Interface (API)
to retrieve the current architecture of the cloud, including the instance
information and security groups. In our architecture we defined a \textit{tag}
named \textit{instance\_type} with the value \textit{mdg} for each instance the
administrator would like to use in constructing Misery Digraph Cloud. Cloud
Constructor identifies those instances and their security groups and leaves
other instances.

By analyzing the rules of security groups, the enabled services in the current
cloud are identified (e.g., web server, FTP server, SSH, etc.).  The Constructor
creates \textit{connectivity digraphs} for each available service. A
connectivity digraph represents the firewall rules that enable communications
amongst two connected instances.  Thus, an edge $(u,v)$ entails that network
communication is enabled from instance $u$ to instance $v$. Based on the created
connectivity digraphs, a misery digraph will be created for each service. In
this step of creating misery digraphs, the network will be filled with decoys
depending on the values of the parameters $d$ and $k$. Once all misery digraphs
of each service are constructed, a union operation will be performed on them to
get the final misery digraph. The final misery digraph is referred to as the
\textit{Misery Digraph Network}.  At this stage, Constructor deploys
\textit{Misery Digraph Network} to the cloud. Depending on the edges of Misery
Digraph Network, security groups and their rules will be created to facilitate
routing network requests across decoy machines in the network. As the new
digraph requires redundant decoy nodes, the corresponding EC2 instances will be
created, resulting in a \textit{Misery Digraph Cloud}.

To enhance the performance, the operating system images of misery Digraph Cloud
should be ready and stored as Amazon Machine Images (AMIs) in the AWS cloud.
Three types of images should be available: (1) A Misery Multicaster image which
will run the entry point and all instances of all layers but layer $d$ (the
layer right before the target) and $d + 1$ (target's layer). (2) Isolated Target
Requests Server image, which runs the layer $d$, web application (that is, an
application that connects to the target service) should reside in this image.
(3) Isolated Target Polling Server image, which runs the target (layer $d+1$),
the database server (which is the attacker's target according to our assumption)
resides in this image.

\subsection{Evolving misery digraphs}

As mentioned earlier in Section~\ref{sec:mdg}, a misery digraph prevents an
intrusion from reaching a target machine by (1) continuously interchanging two
nodes, and (2) deleting and replacing a node with a new node. The goal behind
these two properties of misery digraphs is to lose the effort of an attacker on
an attack path towards the target. The Transformation Process of our
architecture implements these two features. When creating new instances, the TP
faces a challenge: AWS approximately requires five minutes of effort to create a
regular EC2 instance. Since the Transformation Process involves replacing a
running instance with a fresh instance, the five minutes delay causes a
bottleneck.  To mitigate the delays in creating instances, the Movement Manager
(Figure~\ref{fig:ClientRequestComponentOfMDGImplementation}) maintains an
instances pool. The Movement Manager creates a set of $s$ instances in the
instances pool for future use.  Depending on $s$ and the number of running
instances and their image types, the Movement Manager computes the minimum
number of instances needed in the pool while taking into consideration that
creating a new instance during the reset process should occur as infrequently as
possible.
After initializing the instances pool, two random instances will be selected, at
a layer $a$ of the misery digraph, from a random layer but target and entry
point's layers. In other words, the entry point and the target instances shall
not be selected.  The Transformation Process performs two functions: 

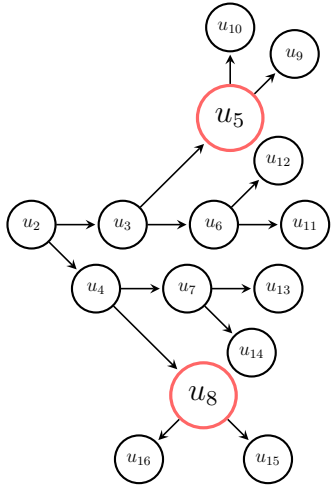
\begin{figure}
\centering
    \begin{tikzpicture}[
            > = stealth, 
            shorten > = 1pt, 
            auto,
            node distance = 2cm, 
            semithick, 
            targetnode/.style={circle, draw=red!60, fill=white!5, very thick, minimum size=2mm},
            entrynode/.style={circle, draw=blue!60, fill=white!5, very thick, minimum size=2mm},
        ]

        \tikzstyle{every state}=[
            draw = black,
            thick,
            fill = white,
            scale=0.6
        ]

        \node[state] (u2) {$u_2$};
        \node[state] (u3) [right of=u2] {$u_3$};
        \node[state] (u4) [below right of=u2] {$u_4$};

        \node[targetnode] (u5) [above right of=u3] {$u_5$};
        \node[state] (u6) [right of=u3] {$u_6$};

        \node[state] (u9) [above right of=u5] {$u_9$};
        \node[state] (u10) [above of=u5] {$u_{10}$};

        \node[state] (u11) [right of=u6] {$u_{11}$};
        \node[state] (u12) [above right of=u6] {$u_{12}$};

        \node[state] (u7) [right of=u4] {$u_7$};
        \node[targetnode] (u8) [below right of=u4] {$u_8$};

        \node[state] (u15) [below right of=u8] {$u_{15}$};
        \node[state] (u16) [below left of=u8] {$u_{16}$};

        \node[state] (u13) [right of=u7] {$u_{13}$};
        \node[state] (u14) [below right of=u7] {$u_{14}$};

        \path[->] (u2) edge node {} (u3);
        \path[->] (u2) edge node {} (u4);
        \path[->] (u3) edge node {} (u5);
        \path[->] (u3) edge node {} (u6);
        \path[->] (u4) edge node {} (u7);
        \path[->] (u4) edge node {} (u8);

        \path[->] (u7) edge node {} (u13);
        \path[->] (u7) edge node {} (u14);
        \path[->] (u8) edge node {} (u15);
        \path[->] (u8) edge node {} (u16);

        \path[->] (u5) edge node {} (u9);
        \path[->] (u5) edge node {} (u10);
        \path[->] (u6) edge node {} (u11);
        \path[->] (u6) edge node {} (u12);

    \end{tikzpicture}
\caption{In this network, the nodes $u_5$ and $u_8$ are randomly selected and switched such that in the new resulting network, $u_8$'s parent is $u_3$, while $u_5$'s parent is $u_4$. }
	\label{fig:MDGSwitchExample}
\end{figure}

\begin{itemize} \item {\it Switching} ensures the disconnection of the
attacker's session, which connects him to a compromised instance from layer $a +
1$. This compromised instance parent is also one of the selected instances. When
this process is performed, the changes will be committed on both Misery Digraph
Cloud in Misery Digraph Network, this process is implemented by switching the
parents and children of the two selected instances. Technically, the security
groups of the two instances will be switched. Figure \ref{fig:MDGSwitchExample}
shows an example of the switching process. The network in this figure is a part
of a complete MDG network. In (A), both $u_5$ and $u_8$ were selected for the
switching process. In (B), the switching was performed, as we can see the parent
and the children of $u_5$ and $u_8$ were swapped.  \item {\it Resetting}
replaces those two selected instances with other pool instances that have the
same image. It ensures that the other instances of the Misery Digraph Cloud can
communicate these new instances. A new pool instance will be created for each
consumed instance pool in the process with the same image type. An instance will
be created on-demand on the case where no pool instance of the image is
available. After finishing this process, the replaced instance will be
terminated. This process helps when the attacker has installed a backdoor or
turned the instance to a bot.  \end{itemize}

\subsection{Multicasting requests}

Misery Digraph requires an instance to multicast any request to the whole
instances in the next layer. It ensures the integrity of the message and hides
the true path to the target.  The Misery Multicaster is a reverse proxy server
with multiple targets instead of just one target. Misery Multicaster runs the
entry point and all instances but layer $d$ instances and the target. 

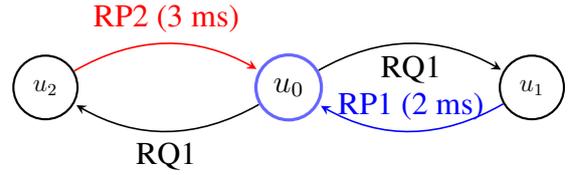
\begin{figure}
\centering
    \begin{tikzpicture}[
            > = stealth, 
            shorten > = 1pt, 
            auto,
            node distance = 4cm, 
            semithick, 
            targetnode/.style={circle, draw=red!60, fill=white!5, very thick, minimum size=2mm},
            entrynode/.style={circle, draw=blue!60, fill=white!5, very thick, minimum size=2mm},
        ]

        \tikzstyle{every state}=[
            draw = black,
            thick,
            fill = white,
            scale=0.8
        ]

        \node[entrynode] (u0) {$u_0$};
        \node[state] (u1) [right of=u0] {$u_1$};
        \node[state] (u2) [left of=u0] {$u_2$};

        \path[->] (u0) edge [bend left, below] node {RQ1} (u1);
        \path[->] (u0) edge [bend left, below] node {RQ1} (u2);
        \path[->][blue] (u1) edge [bend left, above] node {RP1 (2 ms)} (u0);
        \path[->][red] (u2) edge [bend left, above] node {RP2 (3 ms)} (u0);

    \end{tikzpicture}
\caption{When sending the request RQ1 to decoys $u_1$ and $u_2$, the Multicaster awaits for the fastest response, RP1 in this case, and discards the slower ones.}
 \label{fig:MulticasterRequestsAndResponses}
\end{figure}

The first step of a client's communication starts with the misery Multicaster.
We have implemented this component by using an open-source project MapProxy,
which uses the Tornado network framework.  As shown in Figure
\ref{fig:MulticasterRequestsAndResponses}, when a new request is received by
Misery Multicaster, the same request is retransmitted to the instances of the
next layer. Let $ep$ be the entry point, which resides in the first layer and
$u_1$ and $u_2$ be the instances of layer 2. In this case any new request will
be sent to both $u_1$ and $u_2$ by $ep$. Subsequently, $ep$ is going to wait for
the response. Once the fastest instance sends the response back to the $ep$, the
response will be sent directly to the client. The response of the slower
instance(s) will be discarded. The same method is used with a larger number of
instances in the next layer of Misery Multicaster instance.

\subsection{Design of Isolated Target}

The idea of Isolated Target is to isolate the target from any external or
internal connections. Thus, no entity can connect to the target using any port.
While Misery Multicaster is concerned with the client's HTTP request, Isolated
Target addresses the web application's request to the database. As mentioned
earlier, the Isolated Target consists of two components: (1) Requests Server
(RS) and (2) Polling Server (PS). The details of both components will be
examined in the next subsection.

\subsubsection{Requests Server and Polling Server}

In our design, the web application interacts with the \textit{the database
server} as the Requests Server. We developed the Requests Server as a MySQL
server that runs on port 3306 in our setup. Note that the Requests Server is not
a real database management system; it only behaves as one in the handshaking
stage of the connection with the web application. This is completed with one
goal: to get the request that should be sent to the real database and store
those requests.

When a client requests a page from the web application, the request is
transmitted through the network from the entry point by using the Misery
Multicaster. Layer $d$ of the Misery Digraph Cloud contains instances that run
web servers (e.g., Apache HTTPd). The web servers process the request according
to the web application's code. When the code needs to connect to the database
(which is the target), it connects to the  Requests Server instead. 

At the beginning of the session between the web application's code and Requests
Server, the latter is going to behave as a MySQL server to get the request of
the application (e.g., query some table). Once the application's request is
received, it will be assigned with a unique identifier and then be stored in a
location in the Requests Server.  One such location can be the Requests Server's
memory, but we have chosen to store those requests in the disk by using the
SQLite3 database engine. Up to this point, the application is waiting for the
response from what it believes is the database server, and the Requests Server
is waiting for Polling Server.

The Polling Server runs on the target itself. The PS in the same machine the
real database server exists. For each short time interval of $m$, Polling Server
asks all Requests Servers of layer $d$ if there is any new request for the
database that has not been handled. If so, those requests will be consumed by PS
and sent to the real database. Once the response arrives, it will be sent back
to the Requests Server that issued that request. 

\subsubsection{Handling dynamic IP addresses} 
\begin{figure}[ht] \includegraphics[scale=0.28]{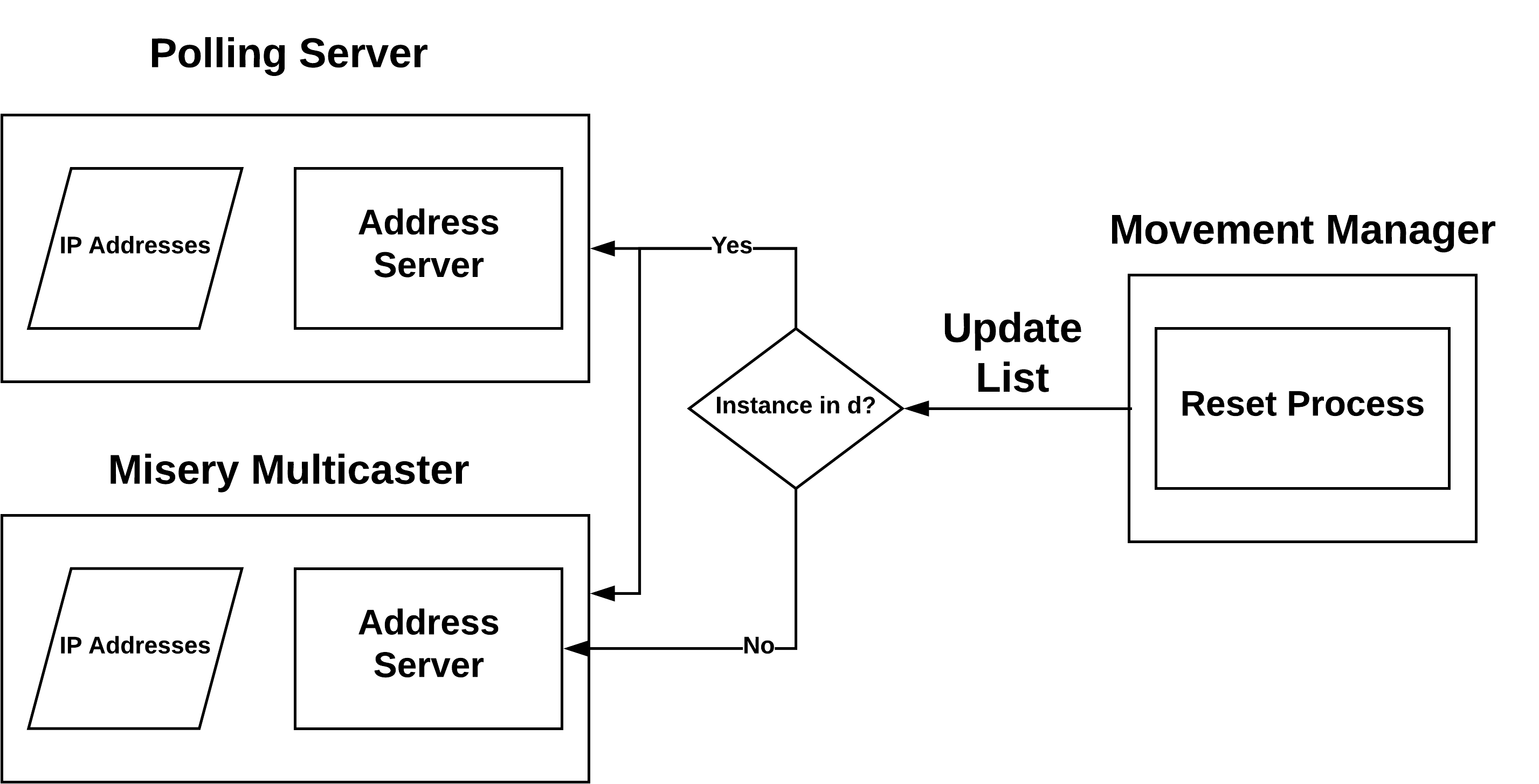} 
\centering \caption{The communication between Movement Manager and Address
Server in both the Misery Multicaster and the Polling Server. When the selected
instances are in layer $d$ the Movement Manager communicates both the Polling
Server's and  the Misery Multicaster's Address Server. Otherwise, the Movement
Manager only communicates with the Misery Multicaster's Address Server.}
\label{fig:AddressServerCommunication} \end{figure}

Recall that Misery Digraph Cloud
evolves over time. When two nodes $u$ and $v$ are interchanged, Misery
Multicaster must update the IP addresses of $u$ and $v$ in the nodes of the next
layer. Similarly, the Polling Server must be updated with the IP addresses of
layer $d$ instances to be able to query them about new requests. Thus, there is
a need to dynamically and efficiently update IP addresses across the misery
digraph. This dynamic updating is handled by the Address Server
(Figure~\ref{fig:AddressServerCommunication}). 

Figure \ref{fig:AddressServerCommunication} shows the communication between the
Movement Manager and Address Server in both the Misery Multicater and the
Polling Server.  Initially, when new instances of the Misery Multicaster and the
Polling Server are started, a list of needed IP addresses will be stored in a
list that will be maintained and used in the future. AWS APIs are to be used to
initialize this list.  When interchanging two instances on any layer (except
layer $d$), the Movement Manager connects to the instances' parent Address
Server to update the Misery Multicaster with the changes. This causes the misery
Multicaster to query AWS API again to receive the new list of next layer's IP
addresses. Similarly, when an instance of layer $d$ is reset. In addition, it
tells its parent to update the IP addresses.  The Address Server of the target
will then be connected by the Movement Manager, and it will be updated with the
new IP addresses of layer $d$. 

%% file: 4.tex
\section{Performance} The main goal of this work is to assess the feasibility of
implementing misery digraphs in real-world networks in terms of the processed
traffic in regular web applications.  For this assessment, multiple rounds of
tests were performed on a synthetic cloud-based virtual network created using an
AWS account. The traffic processed by a normal cloud-based virtual network is
compared with two variations of misery digraphs. The experiments reveal that
while misery digraphs can incur performance penalties  and request processing
failures, the system's performance penalties are reasonable compared to the
magnitude of confusion created for the attacker. 

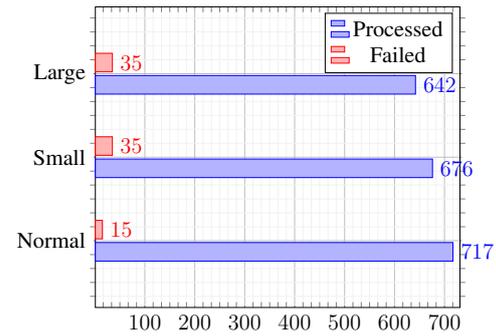
\begin{figure} \centering \begin{tikzpicture}[scale=0.7] \begin{axis}[title  =
Comparison of performance in three cloud-based networks, xbar, y axis line style
= { opacity = 1 }, x axis line style = { opacity = 1 }, tickwidth         = 1pt,
minor tick num=5, grid=both, grid style={line width=.1pt, draw=gray!10}, major
grid style={line width=.2pt,draw=gray!50}, enlarge y limits  = 0.4, enlarge x
limits  = 0.02, symbolic y coords = {Normal,Small,Large}, nodes near coords, ]
  \addplot coordinates{ (642,Large) (717,Normal) (676,Small) }; \addplot
  coordinates{ (35,Large) (15,Normal) (35,Small) }; \legend{Processed, Failed}
  \end{axis}

\end{tikzpicture} \caption{An overview of the results of the experiments with a
normal cloud, a small misery digraph with four layers, and a larger misery
digraph with five layers. } \label{fig:results} \end{figure}

{\it Experimental Setup.} The experiments were performed on t2.nano and t2.micro
instances with a single core processes and 500MB and 1GB RAM, respectively. A
Transformation Process ran on an EC2 instance, performing both switching and
resetting of instances throughout the misery digraph. A request emulation tool
was developed to send a new request to the entry point every few seconds
awaiting the responses.  When a response was received within a timeout window,
the request will be considered a success, otherwise it will be considered a
failure. Recall that $d$ refers to the  number of layers in the Misery Digraph
Cloud, and $k$ refers to the number of children for each instance in misery
digraph network. Let $j$ be the duration of an experiment, $r$ the frequency of
the Misery Process (the number of times the misery digraph changes), and $u$ is
the waiting time for a response. 

The results of the experiments are depicted in Figure~ \ref{fig:results}.
Processed requests capture the number of requests processed by the web
application during the course of an experiment.  The failed requests captures
the number of requests that could not be processed within the same period. The
y-axis values refer to the type of network that was used in the test (cloud with
a small misery digraph, cloud with a large misery digraph, normal cloud with no
misery digraph).

{\it A normal cloud} is one that does not include a misery digraph.  We
conducted an experiment on a normal network which had three EC2 instances. The
first one was an entry point which works as a reverse proxy for the second
instance which ran Apache HTTPd and a tiny PHP page that sent a query to the
third instance which contains a MySQL server. The entry point instance ran
Apache HTTPd with mod\_proxy extension to behave as a reverse proxy server.

All of them were \textit{t2.nano} but the database server was \textit{t2.micro}.
Each instance had 8GB of storage. The operating system images that were used for
the instances were based on Amazon Linux AMI 64-bit. In this experiment $d = 0$
and $k = 0$ since it was a normal network and not a misery digraph cloud, $j =
10$ minutes, $r = 100$ seconds and $u = 1$ seconds.  The goal was to find the
number of requests that normal network could handle and compare those numbers
with networks that includes a Misery Digraph. Our normal cloud could handle an
average of 716.5 requests with a maximum of 15 failed requests.

{\it A Misery Digraph Cloud} extends the architecture of the normal cloud by
including a tree of redundant virtual machines that mediate the entry point and
the target.  We created two Misery Digraph Clouds,  one with $d=3$, and one with
$d=4$, with the branching parameter $k = 2$. We prepared two different operating
system images to be used in the instances pool. The first one contained the
Misery Multicaster based on Amazon Linux AMI, The second one contained the RS
that implemented the Apache HTTPd and a PHP script that connected to a database
for testing using a Ubuntu Server 16.04. Another image was built based on Amazon
Linux AMI, which was used for the Network Constructor that created the Misery
Digraph Cloud, which contained the Polling Server and the target database server
(with a MySQL engine).

In both experiments, $j = 10$ minutes, $r = 100$ seconds and $u = 1.5$ seconds.
As one can see, the average of successful requests with $d=3$ was 675.8, with a
performance penalty of 40.7 requests compared to the normal cloud. When the
misery digraph was created using $d=4$, the average of successful requests was
642 with a performance penalty of 74.5 requests from the normal network. The
most failures were with the executions that performed the Misery Process on the
second layer instances.

%% file: 5.tex
\section{Discussion} Our architecture demonstrates techniques for realizing
misery digraphs in cloud-based virtual networks running conventional web
applications.  Here, we consider both the security and the scalability issues
and provide an analysis and a direction of future work.

{\it The security} of the presented architecture depends entirely on the
security promises of misery digraphs. Attackers are assumed to be remote. The
requests are assumed to be first attack attempts to exploit the target machines
to prepare for malicious data requests to follow through the network. Our
architecture does not distinguish between the two types of attack request and
delays both types regardless of  intents. However, attackers may attempt to
mimic a normal request, taking  advantage of the fair treatment of requests by
misery digraphs as they travel through the network. This attack can be
beneficial only when the initial attacking requests succeed in exploiting
machines in the very first layer of misery digraphs. Consequently, the attacker
must escalate privileges to {\it initiate} new requests from an exploited
machine. This attempt can be prevented by modifying the web application and
banning it from initiating new requests, unless the requests come from a remote
and registered client.

A second possibility of attack is the threat from insiders, which are those
users within the organization that have access to a subset of nodes in the
expanded cloud-based virtual network with an enabled misery digraph. Given
enough nodes en route to the target, the insider may attempt to create malicious
requests or launch an attack on other machines. This is a  vulnerability in the
existing architecture and requires mitigation, which is left for a future work.

{\it The scalability} of the architecture depends on a fast processing and cache
management of misery digraphs. As demonstrated in our work, given current web
server performances, one can create highly scalable misery digraphs with low
error rates. One may suggest creating even larger misery digraphs for better
mitigation of the attack. However, as demonstrated in~\cite{8125739}, a misery
digraph with $d=4$ with fast switching is confusing enough for attackers. It
remains a question whether misery digraphs are effective against distributed
attacks, another topic for future investigation.

%% file: 6.tex
\section{Conclusion} Prior to this work, the practical and performance
feasibility of misery digraphs were not systematically explored. This research
contributes to the idea of misery digraphs by presenting an efficient and high
throughput architecture that can be used in practice. The main constraints of
this work were to use existing prominent cloud technologies and high request
processing performance, which were achieved. In the future, this work will be
expanded to explore the idea of adaptive changes to the underlying misery
digraph as a real time response to detected attack incidents.